\documentclass[12pt]{article}

\def\spose#1{\hbox to 0pt{#1\hss}}
\newcommand{\en}{\end{displaymath}}
\newcommand{\bea}{\begin{eqnarray}}
\newcommand{\ena}{\end{eqnarray}}
\newcommand{\e}{\mbox{e}}

\def\past{\mathrm{past}}
\def\future{\mathrm{future}}
\def\maximal{\mathrm{maximal}}
\def\singlespace{\baselineskip=12pt}      
\def\sesquispace{\baselineskip=16pt}      

 at10pt

\def\past   {\mathop {\rm past     }\nolimits}
\def\future {\mathop {\rm future   }\nolimits}
\def\maximal{\mathop {\rm maximal\,}\nolimits}

\def\=>{\Rightarrow}
\def\==>{\Longrightarrow}

 \def\dal{\displaystyle{{\hbox to 0pt{$\sqcup$\hss}}\sqcap}}

\def\lto{\mathop
        {\hbox{${\lower3.8pt\hbox{$<$}}\atop{\raise0.2pt\hbox{$\sim$}}$}}}
\def\gto{\mathop
        {\hbox{${\lower3.8pt\hbox{$>$}}\atop{\raise0.2pt\hbox{$\sim$}}$}}}
\def\half{{1\over 2}}

\def\less{\backslash}		

\def\to{\mathop\rightarrow}	

\def\tilde{\widetilde}		

\def\ideq{\equiv}		

\def\interior #1 {  \buildrel\circ\over  #1}     

\def\basisvector#1#2#3{
 \lower6pt\hbox{
  ${\buildrel{\displaystyle #1}\over{\scriptscriptstyle(#2)}}$}^#3}

\def\eprint #1 {{$\langle$e-print archive: #1$\rangle$}}

\usepackage{latexsym}
\usepackage[dvips]{graphics,color}
\def\be{\begin{displaymath}}
\def\ee{\end{displaymath}}
\def\bne{\begin{equation}}
\def\ene{\end{equation}}
\def\bee{\begin{eqnarray*}}
\def\eee{\end{eqnarray*}}

\def\P{M}			
\def\vpi{\varpi}
\def\eps{\epsilon}

\begin{document}


\phantom{}
\vskip -.8 true in 
\medskip
\rightline{gr-qc/0009063}
\rightline{SU--GP--00/7--1}
\vskip 0.3 true in


\bigskip
\bigskip


\sesquispace
\large
\centerline{\bf On the ``renormalization'' transformations induced by}
\centerline {\bf cycles of expansion and contraction in causal set cosmology}
\bigskip


\singlespace			        
\normalsize
\centerline {\it Xavier Mart{\'\i}n$^{(1)}$}
\centerline {\it Denjoe O'Connor$^{(2)}$}
\centerline {\it David Rideout$^{(3)}$}
\centerline {\it Rafael D. Sorkin$^{(4)}$}

\medskip
\smallskip
\centerline {\it $^{(1,2)}$Depto. de F\'{\i}sica, CINVESTAV, 
Apartado Postal 70543, M\'exico D.F. 07300, MEXICO}
\smallskip
\centerline {\it $^{(3,4)}$Department of Physics, 
                 Syracuse University, 
                 Syracuse, NY 13244-1130, U.S.A.}
\smallskip
  
\centerline {\it \qquad\qquad internet addresses: $^{(1)}$xavier@fis.cinvestav.mx, $^{(2)}$denjoe@sirius.fis.cinvestav.mx,}
\centerline{\it \quad\quad $^{(3)}$rideout@physics.syr.edu, $^{(4)}$sorkin@physics.syr.edu}

\begin{abstract}
We study the ``renormalization group action'' induced by cycles of
cosmic expansion and contraction, within the context of a family of
stochastic dynamical laws for causal sets derived earlier.  We find a
line of fixed points corresponding to the dynamics of transitive
percolation, and we prove that there exist no other fixed points and
no cycles of length $\ge{2}$.  We also identify an extensive ``basin
of attraction'' of the fixed points but find that it does not exhaust
the full parameter space.  Nevertheless, we conjecture that every
trajectory is drawn toward the fixed point set in a suitably weakened
sense.
\end{abstract}


\sesquispace

\section {Introduction} 
There is good reason to believe that the appropriate dynamical framework
for ``quantum gravity'' will be --- just as the ordinary quantum
dynamical framework already is --- a generalization of classical
probability theory.  If in addition, we take the deep structure of
spacetime to be that of a causal set \cite{gcr}, 
then the required dynamics
will be expressible as a generalized probability measure on the space of
all causal sets, the generalization being in the sense of \cite{qmqmt} 
and \cite{gqmd}
or something similar.

In the search for such a dynamics, principles like
general covariance and relativistic causality can offer much needed
guidance.  Indeed, in the limit in which quantal probabilities reduce to
classical ones, it has proved possible to derive \cite{csgd}
from versions of these
two principles a unique family of dynamical laws (Markov processes)
parameterized by a sequence of ``coupling constants''
\bne
  t_0, \ t_1, \ t_2, \  \cdots \,,
\label{D3}
\ene
where $t_0\ideq{1}$ and $t_n\ge{0}$ for $n=1,2,3,\cdots$.  Although this
family of stochastic processes is tremendously special compared to the
most general relevant Markov process, it still contains a countable
infinity of coupling constants, and the question arises how further to
narrow down the field of possibilities.  Any answer to this question can
be expected to be of use in the attempt to find a quantal generalization
of the dynamical scheme in question; it might also, with luck, lead us to
a choice of the $t_n$ which would define a causal set version of
classical general relativity, arising as the classical limit of quantum
causet dynamics. 
(We will often use the contraction ``causet'' in place of the longer
``causal set''.)

In the absence of any further general principle\footnote{
One principle that suggests itself is that of ``locality''.
Unfortunately, this does not seem to have meaning (at a fundamental
level) for causets, though it must, of course, emerge in a suitable
continuum approximation if the theory is to reproduce low energy
physics correctly.}
that might be brought to bear in narrowing down the possibilities
(\ref{D3}), one can consider an ``evolutionary'' approach to this
problem, that is, one can try to let the causet ``choose its own
dynamics''.  Such a possibility was discussed in \cite{peyresq}, where
it was shown that each cosmic cycle of expansion and contraction for
the causet has the effect of ``renormalizing'' the parameters $t_n$ in
a definite manner.\footnote
{The question of which dynamical laws actually lead to such cycles (as opposed to monotonic expansion, for example) is very much open.  Only in the case of percolation is it known that an infinite number of cycles occur with probability 1.}
 The renormalization reduces, in fact, to the
repeated action of a single transformation $\P$, given by equation
(\ref{D9}) below.  By understanding better the action of $\P$, one
might hope to identify a distinguished sub-family of the dynamical
laws parameterized by (\ref{D3}).  For example, if $\P$ had a single
fixed point $(t^0_n)$ and if every other set of parameters $(t_n)$
were carried toward $(t^0_n)$ by the action of $\P$, then one might
designate $(t^0_n)$, or more generally some neighborhood thereof, as
distinguished since it would be the dynamics toward which the universe
would naturally evolve.\footnote{
In much the same manner, one has explained why low energy physics ``must''
be described by a {\it renormalizable} quantum field theory.  In that
case, however, the RG flow in question is conceptual, being an evolution
in energy scale, whereas here it is truly a change with time.  (Of
course, evolution in time gets correlated with evolution in
energy/length scale in the context of big bang cosmology.) }

We will see below that the actual situation is more complicated than
this, but with many points of similarity.  Instead of a single fixed
point, there is a one-parameter family of them, and not all sequences
$(t_n)$ are attracted to the fixed point set, at least in the most
obvious sense.  We will also find that the transformation $\P$ has no
``limit cycles'', but we do not know whether it might have
``attractors'' of a more complicated sort.  Thus, we will make a start
on determining the nature of the ``RG flow'' associated with $\P$, but
we will not characterize it fully.  Nor will we obtain much
information about the flow in the neighborhood of the fixed point set
itself.  Some such results for specific trajectories can be found in
\cite{dou} and \cite{peyresq}, however.  Indeed, the very suggestive
nature of some of those results for helping to explain the most
salient features of the early universe furnished a good part of the
motivation for the present investigation.

\section{Formulation of the problem and summary of our main results}
\label{formulation}
Let us now pose more precisely the problem we intend to address.  In
the course of the growth/evo\-lu\-tion of a causal set $C$, a {\it post}
occurs when an element $\omega$ is born such that every other element
of $C$ is (or will be) either an ancestor or a descendant of $\omega$.
That is, a post is an element $\omega$ of $C$ for which
$$
    C = (\past \, \omega) \cup  \{ \omega \}  \cup (\future\,\omega) \,,
$$
where
$\past(\omega) := \{x\in C | x \prec \omega \}$
and
$\future(\omega) := \{x\in C | \omega \prec x \}$ 
(Our basic definitions and notation are as in \cite{csgd}, in
particular $\prec$ is irreflexive.)  Cosmologically speaking, this may
be interpreted as a kind of collapse of the universe to zero diameter
followed by a subsequent re-expansion.  It is a simple consequence of
the form taken by the dynamical law derived in \cite{csgd} that the
portion
$\{\omega\}\cup(\future\,\omega)$
of $C$ comprising the post and its descendants is governed by an
effective dynamics of the same nature as that valid for $C$ as a whole,
but with different values of the $t_n$.

To see what these new parameters are, recall from \cite{csgd}
that any given transition probability leading from an $n$-element causet
to one of $n+1$ elements takes the form 
\bne
     { \lambda(\vpi,m) \over \lambda(n,0) } \,,
\label{D8-1}
\ene
where
\bne
    \lambda(\vpi,m) := \sum\limits_{k=m}^\vpi 
    {\vpi - m \choose k-m}  t_k  \,.
\label{D8-2}
\ene
Here the transition is occasioned by the birth of some new element $x$
and $\vpi=|\past(x)|$ is the number of all its ancestors, while
$m=|\maximal(\past(x))|$ is the number of its {\it immediate}
ancestors or ``parents''.  Now, by definition, no element $x$ born to
the future of a post can have vanishing $m$ (or $\vpi$).  In view of
(\ref{D8-2}), it follows simply from this that the parameter $t_0$
becomes irrelevant subsequent to a post.\footnote
{In more detail, the reasoning is that the probabilities of the
possible transitions (those that respect $\omega$ being a post) are,
according to (\ref{D8-1}), given by the $\lambda(\vpi,m)$, up to an
overall normalization, and since $m=0$ is excluded, $t_0$ never occurs
in these, according to (\ref{D8-2}).  Since the normalization then
follows from the requirement that all allowed probabilities sum to 1,
it also cannot depend on $t_0$.  This is made explicit in
(\ref{D11-1}) below (see also the Appendix).}
By the same token, we see from (\ref{D8-1}) that only the ratios of
the remaining $t$'s, $(t_1,t_2,t_3,\cdots)$, have meaning in such a
region of $C$, not their overall normalization.  Moreover, if $\vpi$
and $m$ are the {\it effective} sizes of $\past(x)$ and of
$\maximal(\past(x))$ for some element $x$ born after the post
$\omega$, then their true values are plainly $\vpi+p$ and $m$,
respectively, where $p=|\past\omega|$ is the number of ancestors of
the post.  The (un-normalized) probability of this birth is therefore
given by the effective quantity $\tilde\lambda(\vpi,m)$, where
\bne
   \tilde\lambda(\vpi,m) = \lambda(\vpi + p, m)  \,.
\label{tlam}
\ene

Given these facts, 
it is an easy matter to derive 
the effective coupling constants for the after-post dynamics, 
and we claim 
that $\tilde{t}^{\,(p)}_n$, 
the effective or ``renormalized'' value of $t_n$ in the after post region, 
is given by
\bne
  \tilde{t}^{\,(p)}_n = \sum\limits_{k=0}^p {p \choose k} t_{n+k} \,, 
\label{D9-1}
\ene
or equivalently by $p$ applications of the transformation $\P$ to the 
fundamental coupling constants, where $p=|\past\omega|$ 
and $\P$ takes $(t_1,t_2,t_3,\cdots)$ 
to $(\tilde{t}_1,\tilde{t}_2,\tilde{t}_3,\cdots)$  via
\bne
   \tilde{t}_n = t_n + t_{n+1} \qquad (n=1,2,3,\cdots) \,.
\label{D9}
\ene

Perhaps the simplest way to demonstrate these claims, if we accept that
a relationship of the form (\ref{D8-2}) holds good for the effective
dynamics\footnote{
Essentially, this is proven in \cite{csgd}, however in strict logic, one
would have to redo the whole derivation for the present ``originary''
case, where no new element is born with an empty past, this being
part of the meaning of a post.},
is just to observe that, 
as a trivial special case of that equation,
$$
   t_n = \lambda(n,n)
$$
from which we find immediately, for the special case $p=1$,
$$
    \tilde{t}_n = \tilde\lambda(n,n) = \lambda(n+1,n) = t_n+t_{n+1} \,.
$$
Our claim that the $\tilde{t}^{\,(p)}_n$ are given by $p$ applications
of $\P$
is then a consequence of the manifest fact that adding $p>1$ elements
to $\past(\omega)$ all at once is no different from adding them one
by one. 
A logically impeccable proof is almost as easy, but requires a
short calculation, given in the Appendix, to establish (\ref{A1}).
In the light of the latter, one sees immediately from (\ref{tlam}) that the
effective coupling constants are indeed those of (\ref{D9-1}), and these in
turn are readily seen to be given by $\P^p(t_n)$, for example by
noting that $\P$ can be expressed as $1+a$, where $a$ is a shift
operator acting on the $t_n$ and comparing (\ref{tlam}) with the binomial
expansion of $(1+a)^p$.

The ``renormalization (semi)group'' we will study
in the sequel is that generated by the transformation $\P$.  For
consistency with $\omega$ being a post, at least one of the parameters
$t_1,t_2,t_3\cdots{}t_{p+1}$ must be nonzero.  Henceforth we will assume for
simplicity that $t_1>0$. If this is not true initially, it will become
so after a sufficient number of applications of the operator $\P$.

For completeness, we record here the correctly normalized transition
probability $T$ for the birth of an element $x$ to the future of our
post $\omega$.  For this purpose, let 
$\varpi=|\past x|-p$ be the {\it effective} number of ancestors of $x$
(excluding the $p$ elements preceding $\omega$), let $m$ be its number
of parents, and let $n$ be the effective size of the pre-existing causet
(the size of $C\less(\past\omega)$ before $x$ is born).
Then $T$ can be written as
\bne
    { \tilde\lambda(\vpi,m) \over \tilde\lambda(n,0)-\tilde\lambda(0,0) }
\label{D11-1}
\ene
where $\tilde\lambda(\vpi,m)$ is the same function of the
$\tilde{t}_k$ as 
$\lambda(\vpi,m)$ is of the $t_k$:
$$
  \tilde\lambda(\vpi,m) = \sum\limits_{k=m}^\vpi
  {\vpi-m\choose k-m} \tilde{t}_k
  = \lambda(\vpi+p,m) \,,
$$
and where, consequently, the denominator can be written more properly as
$$
  \sum\limits_{k=1}^n {n \choose k} \tilde{t}_k 
$$
(eliminating the apparent reference to $\tilde{t}_0$).

In the next section, we will prove those properties of the
``renormalization map'' $\P:t_n\to{}\tilde{t}_n$ that we have been able
to establish.  We assume throughout that $(t_n)=(t_1,t_2,t_3,\cdots)$ is
a sequence of nonnegative real numbers, with $t_1>0$, and we let $\P$
act by (\ref{D9}), it being understood that the space ${\cal T}$ on which
it acts is actually the set of equivalence classes of sequences $(t_n)$,
where $(t_n)$ and $(t'_n)$ are equivalent iff $t_n=\lambda{}t'_n$ for
all $n$ and for some fixed $\lambda>0$.
Our principal results are then as follows.

\noindent (i) 
The fixed points of $\P$ are given by the sequences
$(t_n)$ such that 
\bne
              t_n = t^n     \label{D12}
\ene
for some $t\ge{0}$ \cite{dou}.  
They thus form a 1-parameter set, whose parameter
$t$ is related to the parameter $p$ of originary percolation \cite{csgd}
by $p=t/(t+1)$.  In (\ref{D12}) the $t=0$ case is to be interpreted by
taking the limit $t\searrow{0}$,
which is equivalent to putting $t_n=\delta_{n1}$, a dynamics which
produces originary causets $C$ that are always trees \cite{peyresq,csgd}.
Note that the limit $t\to\infty$ also makes sense, and corresponds to
originary percolation with $p=1$ (cf. (\ref{D8-2}) and (\ref{D11-1})), 
a dynamics  which always produces the same causet: the ``purely one
dimensional'' poset, or chain.

\noindent (ii)
Aside from its fixed points, $\P$ possess no other cycles. 

\noindent (iii)
Suppose that $(t_n)$, though not necessarily of the form (\ref{D12}), is
such that $t_n^{\,1/n}$ has a limit in $[0,\infty)$
as $n\to\infty$.  
Then, under
repeated action of $\P$, the sequence $(t_n)$ converges pointwise to
(\ref{D12}) with $t=\lim\limits_{n\to\infty}t_n^{\,1/n}$.  
(Of course, 
this convergence can only be modulo 
the overall scale ambiguity in $(t_n)$.
One way to lift this ambiguity is 
to deal with the {\it ratios} $t_{n+1}/t_n$,
and what we prove below is that
$$
 \lim\limits_{p\to\infty}\tilde{t}^{\,(p)}_{n+1}/\tilde{t}^{\,(p)}_{n}=t \,,
$$
where $(\tilde{t}^{\,(p)}_{n})$
is the result of acting $p$ times with $\P$ on $(t_n)$.  
Notice that
this asserts more than simply pointwise convergence to $t_n=\delta_{n1}$
in the case $t=0$.
Notice also that we have not included the case $t=\infty$ in the result
just stated.)

For the case where $t_n^{\,1/n}$ has no $n\to\infty$ limit, we have no
general result, although one can show 
using the generating functions defined below 
that
if $(\tilde{t}^{\,(p)}_n)$ {\it does} converge pointwise 
to some fixed point (\ref{D12}), 
then 
the latter must be the one with parameter $t=\limsup{t_n^{\,1/n}}$.  
In general, however, one can not expect
pointwise convergence to any sequence 
(see the counterexample in Section \ref{counterex} below.)

As a matrix, our transformation $\P$ is just 
$$
    \pmatrix{1&1&0&0&0&\cdots \cr
             0&1&1&0&0&\cdots \cr
             0&0&1&1&0&\cdots \cr
             \vdots&\vdots&\vdots&\vdots&\vdots&  \cr}
$$
Given the simplicity of this matrix (it is already in ``Jordan normal
form'' and in fact is just the identity plus a shift operator), 
one might think it an easy matter to understand the
``flow'' its powers define.  However one meets with two complicating
circumstances: the infinite dimensionality and the fact that $\P$ only
{\it appears} to be linear, because the space ${\cal T}$ on which it acts
is really a projective space (rather than a vector space), since its
points are given by the {\it ratios}, 
$t_1:t_2:t_3\cdots$.  For these reasons, we have chosen to study the
transformation (\ref{D9}) directly, without attempting to utilize any of
the general results (such as the spectral theorem, for example) which
one might have tried to bring to bear on the problem.

\section{Functional representation}

We are working on sequences $t_n$ ($n=1,2,3,\cdots$) 
defined up to an overall multiplicative constant, 
and such that 
\bne
     t_n\geq 0, \  t_1\not= 0 \,.  \label{cond} 
\ene
One way to remove the overall multiplicative constant freedom is to
set $t_1=1$, however it will be more convenient to keep the freedom in
the following. The renormalization scheme on these sequences is
given by 
\bne 
    \tilde{t}_n=t_n+t_{n+1}.\label{renorm1} 
\ene
It is rather clear that after renormalization, the new sequence
satisfies the same conditions (\ref{cond}) as the initial
sequence.  Thus it is possible to iterate the renormalization, and
the issue is then to decide what kind of limiting behavior arises
after a large number of iterations.

It is sometimes useful to represent these sequences as formal power
series defined up to a multiplicative constant: 
\be
    G(z)=\sum_{n=1}^\infty t_nz^n. 
\en
The sequence can be recovered from the power series by differentiation:
\be
   t_n=\frac{G^{(n)}(0)}{n!} \,.
\en
In this new representation, the renormalization (\ref{renorm1}) is
mapped to a functional relation between power series which can be
derived easily as 
\bea
  \tilde{G}(z) & = & \sum_{n=1}^\infty \tilde{t}_n z^n=\sum_{n=1}^\infty
  t_nz^n+\sum_{n=2}^\infty t_nz^{n-1} \nonumber \\
  & = & (1+\frac{1}{z})G(z)-G'(0).\label{renorm2} 
\ena
Iterating this functional relation is still rather difficult, however
a change of variable 
\bee
  \frac{1}{y} & = & 1+\frac{1}{z},\\
  g(y) & = &  G(z)
\eee
induces a new renormalization 
map 
on $g(y)$ which is simply given by
\bne
   \tilde{g}(y)=\frac{g(y)}{y}-g'(0) \,. \label{renorm3} 
\ene
Note that since the change of variable 
is analytic, $g(y)$ is still a formal power series 
\be
    g(y)=\sum_{n=1}^\infty g_n y^n
\en
with new coefficients which depend on the 
original ones
as 
\be
   g_n=\sum_{p=1}^{n} {n-1 \choose p-1} t_{p}  \,.
\en
There is no obvious sufficient condition on the $g_p$ which will
ensure the positivity of the $t_n$, so that this condition can not be
checked easily in the $y$ representation. 
However, it is easy to see that it implies that $g_p$ is bounded below
by some strictly positive number (because $g_p\geq t_1>0$).
The renormalization equation (\ref{renorm3}) 
written in terms of the coefficients $g_n$ 
takes a particularly simple form 
\be
  \tilde{g}(y)=\sum_{n=1}^\infty g_{n+1} y^n, 
\en
which can evidently be iterated $p$ times to get 
\be
  \tilde{g}^{(p)}=\sum_{n=1}^\infty g_{n+p} y^n \,.
\en

\section{Fixed points}

Fixed points are sequences $(t_n)$ which do not change under
renormalization. 
Since our sequences are only defined up to a
multiplicative constant, this means that 
\be
     \tilde{t}_n = c t_n
\en
or equivalently
\be
       t_n + t_{n+1} =  c t_n
\en
the solution of which is
\be
        t_{n+1} = (c-1) t_n
\en
where we must have $c\ge{1}$ in consequence of the positivity of the $t_j$.
For $c>1$, this implies $t_n = t^{n-1}t_1$, where we have put $c-1=t$; 
this can be written most compactly if we use the scale freedom to  
set $t_1$ itself to be $t$, 
in which case we get simply
\bne
        t_n = t^n   \,.
\ene
For $c=1$, we just obtain the $c\to{1}$ limit of these relationships, 
namely
$$
    t_1=1, \ t_2=t_3=t_4=\cdots=0 \,.
$$

The equivalent relationships in terms of the generating functions
introduced in the last section are as follows.
Fixed points are power series which do not change under
renormalization. Since these power series are only defined up to a
multiplicative constant, this means that 
\bne
 \tilde{G}(z)=tG(z),\label{fixede} 
\ene
$t$ a real (positive) constant. Putting Eq. (\ref{renorm2}) in
(\ref{fixede}) then gives 
\be
 G(z)\propto\frac{z}{1-tz}.\label{fixed} 
\en
This corresponds for the sequence $(t_n)$ to a geometric series of
ratio $t$, and gives a power series 
\be
    g(y) \propto \frac{y}{1-cy} 
\en 
of the same type after change of variable,
where we have put $t+1=c$.

\section{Cycles}

Cycles are such that, after a finite number of renormalizations $p$, 
one gets back to the initial sequence. Taking into account the
multiplicative constant freedom, this gives the equation 
\bne 
   \tilde{t}^{(p)}_n=c^p t_n,\label{eqcycle} 
\ene
for all $n\geq 1$, where the constant, 
which is necessarily positive 
since both $t_n$ and $\tilde{t}^{\,(p)}_n$ are, was written as
a power $c^p$. 
In light of (\ref{D9-1}), 
the equations (\ref{eqcycle}) can be rewritten as a 
recursion relation 
for the sequence $t_n$ as 
\bne
  t_{n+p}=(c^p-1)t_n-\sum_{q=1}^{p-1} {p \choose q}
  t_{n+q}.\label{recurs} 
\ene
For the sequence to be positive, this implies in particular $c\geq 1$. 
A sequence given by such a linear 
recursion relation 
with constant coefficients independent of the index $n$
can always be rewritten as a linear
combination of $p$ geometric progressions satisfying (\ref{recurs}). 
The corresponding polynomial characteristic equation
for the ratios can be solved easily as
\be 
  q_j=-1+c\e^{2i\pi j/p},
\en
and a cycle must therefore have the general form 
\be
  t_n= \alpha_0 (c-1)^n 
  +
  2\sum_{1\leq j<p/2} \mbox{Re}(\alpha_j q_j^n) \quad
  (+ \  \alpha_{p/2} (-c-1)^n)
\en
where the last term (in parentheses) is only there if 
$p$ is even $\alpha_0$ and $\alpha_{p/2}$ are real constants, the
other $\alpha_j$ are complex, and the combination was chosen such that
$t_n$ be real.

Note that 
$\alpha_{p/2}(-c-1)^n = (-)^n \alpha_{p/2}(c+1)^n$ 
will never be purely positive for $n$
large. Similarly, $\mbox{Re}(\alpha_j q_j^n)$ can never be purely
positive for large $n$. Indeed, if the phase of $q_j$ is called
$\theta_j$, it is well known that for $\theta_j/2\pi$ irrational,
$\cos(n\theta_j)$ is dense in $[-1,1]$, whereas if it is rational, it
is periodic with values of the form $\cos(2k\pi/r)$, $0\leq k <r$,
which is positive for $k=0$ and negative for $k=[r/2]$, the integer
part of $r/2$.

As a linear combination of geometric progressions, the behavior, and
the sign, of $t_n$ for large $n$ is dominated by the geometric
progression with non zero coefficient $\alpha_j$ and ratio $q_j$ with
largest modulus. For $t_n$ to be positive for large $n$, the geometric
progression $(c-1)^n$ must therefore dominate there. As it happens, the
modulus of $q_j$ grows with $j$ from $j=0$ to $j=[p/2]$, which implies
that for $0<j\leq p/2$, all the coefficients $\alpha_j$ must be
$0$. This gives for the sequence $t_n$ the necessary form 
\be
  t_n=\alpha_0 (c-1)^n,
\en
which is 
percolation with $t=c-1\geq 0$, and therefore a fixed point.

Thus, we have proved that the only cycles are the fixed points.

\section{Flows}

The issue is now to determine the behavior of a given initial
sequence after a large number of renormalizations. 
Among other possibilities, one may expect
either convergence to a fixed point, 
or some kind of oscillatory behavior.  
The convergence studied here will simply be 
pointwise convergence of sequences.
In the absence of cycles, 
and since fixed points are geometric progressions (percolation), 
a reasonable hypothesis is that 
an initial sequence will converge to a fixed point 
described by $t=\lim(t_n^{1/n})$,
assuming this limit exists.
In the following, 
this hypothesis will be 
validated in two qualitatively different cases: 
$t>0$ finite, 
and its limiting case, $t=0$.  
If $t_n^{1/n}$ does not have a limit (or converges to $\infty$),
the issue becomes more complex, 
as will be discussed in the last part of this section and subsequently.

\subsection*{t finite and non-zero}
In this subsection, the assumption will be that 
\bne
  \lim_{n\rightarrow\infty}\left( t_n^{1/n} \right) =t\not= 0.
  \label{dalem}
\ene
As a consequence, one can write 
\bne
  t_{n}=t^n (1+\varepsilon_n)^n ,\label{expres} 
\ene
where $\varepsilon_n$ is an auxiliary sequence which goes to $0$ as
$n\rightarrow\infty$.  Another useful auxiliary sequence is 
\be
   E_n=\sup_{i\geq n}(|\varepsilon_i|) 
\en
which also goes to $0$ as $n\rightarrow\infty$.

Taking into account the scaling freedom of the sequences, the
pointwise convergence of the sequence 
to a fixed point
in the limit of a large number
of renormalizations is expressed as 
\be
  \lim_{p\rightarrow +\infty} \,
  \frac {\tilde{t}_n^{(p)}} {\tilde{t}_1^{(p)}} = t^{n-1} \,,
\en 
for $n$ a fixed integer number. 
This is equivalent to showing that for
any given $n\ge{1}$, 
\bne
 \lim_{p\rightarrow +\infty} \,
 \frac{\tilde{t}_{n+1}^{(p)}}{\tilde{t}_n ^{(p)}}  = t \,. 
 \label{limproof} 
\ene

First, it will be useful to prove the following Lemma. 
\paragraph{Lemma} 
If $t_n \geq 0$ is a sequence satisfying (\ref{dalem})  
for some particular $t>0$
and if 
$i_1,i_2,i_3,\cdots$ is a sequence of positive integers
such that 
$i_p=o(p/\ln p)$,
then 
\bne
  \sum_{i=0}^{i_p} \left( \! \begin{array}{c} p\\i\end{array} \! \right)
  t_i
  =
  o \left(
     \sum_{i=0}^p \left( \! \begin{array}{c} p\\i\end{array} \! \right)
     t_i \right) 
  \,.
\label{lemma}
\ene
We will need the lemma only in the special case $i_p=\sqrt{p}$.

The proof will be obtained by showing first that the sequence $t_n$ can be
replaced by another sequence $u_n$ with no zeros so that the sequence
$u_n^{1/n}$ be bounded from below by a strictly positive number. From
(\ref{dalem}), it is clear that an integer $i_0$ can be fixed such
that $t_i>0$ for all $i\geq i_0$. 
Then introducing the auxiliary sequence $u_n$ such that 
\bee
 u_n & = & t_n+1 \mbox{ if } n \le i_0 \\
 u_n & = & t_n \mbox{ otherwise} 
\eee
one gets (when $i_p>i_0$)
\be
 \frac{\sum_{i=0}^{i_p} \left( \! \begin{array}{c} p\\i\end{array}
 \! \right) t_i}{\sum_{i=0}^p \left( \! \begin{array}{c} p\\i
 \end{array} \! \right) t_i} =\frac{\sum_{i=0}^{i_p} \left( \!
 \begin{array}{c} p\\i\end{array} \! \right) u_i-
 \sum_{i=0}^{i_0} \left( \! \begin{array}{c} p\\i\end{array} \! \right)
 }{\sum_{i=0}^p \left( \! \begin{array}{c} p\\i\end{array} \! \right)
 u_i-\sum_{i=0}^{i_0} \left( \! \begin{array}{c} p\\i\end{array}
 \! \right)} \ .
\en 
Moreover, 
\be
 \sum_{i=0}^{i_0} \left( \! \begin{array}{c} p\\i\end{array} \! \right)
 =o(\left( \! \begin{array}{c} p\\i_0+1 \end{array} \! \right) u_{i_0+1})
\en
because the left hand side is a polynomial in $p$ of degree $i_0$
while the right hand side is a polynomial of degree $i_0+1$. 
Thus
(with `$A\sim{}B$' meaning as usual that $A/B\to1$),
\be
\frac{\sum_{i=0}^{i_p} \left( \! \begin{array}{c} p\\i\end{array}
\! \right) t_i}{\sum_{i=0}^p \left( \! \begin{array}{c} p\\i
\end{array} \! \right) t_i} \sim \frac{\sum_{i=0}^{i_p} \left( \!
\begin{array}{c} p\\i\end{array} \! \right) u_i}{\sum_{i=0}^p \left
( \! \begin{array}{c} p\\i \end{array} \! \right) u_i} \en
and the sequence $t_n$ has been replaced by a strictly positive
sequence that fulfills (\ref{lemma}) iff the original sequence does.
So in the following, it will be simply assumed that $t_n$ was strictly
positive to start with. 

Let $0<T_-\leq T_+$ be the lower and upper bounds of the sequence
$t_n^{1/n}$. Then, 
\be 
\frac{\sum_{i=0}^{i_p} \left( \! \begin{array}{c} p\\i\end{array} \!
\right) t_i}{\sum_{i=0}^p \left ( \! \begin{array}{c} p\\i \end{array}
\! \right) t_i} \leq \frac{\sum_{i=0}^{i_p} \left( \!
\begin{array}{c} p\\i\end{array} \! \right) T_+^i}{(1+T_-)^p} .
\en
The numerator can be bounded from above by comparing it with a
geometric series. 
Thus, call 
\be 
v_i=\left( \! \begin{array}{c} p\\i\end{array} \! \right) T_+^i \ ,
\en
then 
\be
\frac{v_i}{v_{i+1}}=\frac{i+1}{p-i}\ \frac{1}{T_+}\leq \frac{i_p+1}{p 
-i_p}\ \frac{1}{T_+}\rightarrow 0
\en
since $i_p=o(p)$. So, by taking $p$ large enough, it can be ensured
that \be
\frac{v_i}{v_{i+1}}\leq\frac{1}{2}.\en
Then, 
\bne
  \frac{\sum_{i=0}^{i_p} \left( \! \begin{array}{c} p\\i\end{array} \!
  \right) t_i}{\sum_{i=0}^p \left ( \! \begin{array}{c} p\\i \end{array}
  \! \right) t_i}\leq \frac{ 2\left( \! \begin{array}{c} p\\i_p
  \end{array} \! \right) T_+^{i_p}}{(1+T_-)^p} \ .
  \label{interlemma} 
\ene
The numerator on the right hand side can be bounded using that
$$
  {p\choose k} = {(p-0)(p-1)(p-2)\cdots(p-[k-1])\over k!}
               \le {p^k \over k!}  \ ,
$$
where we have put $i_p=k$ for short.  From this it follows that
$$
 {2{p\choose k}{T_+}^k\over(1+T_-)^p} \le { 2(pT_+)^k \over (1+T_-)^p }
$$
or taking logarithms,
$$
\ln{2{p\choose k}{T_+}^k\over(1+T_-)^p}
\le 
\ln 2 + k \ln(pT_+) - p \ln(1+T_-)
\sim
-p \ln(1+T_-)
\to -\infty \,
$$
which implies in turn that the right hand side of (\ref{interlemma})
converges to zero for $p\rightarrow\infty$ and concludes the proof of
the Lemma.

\paragraph{Main proof}
The desired limiting behavior, equation (\ref{limproof}), can be rearranged
using Eqs. (\ref{D9-1}) and (\ref{expres}) as 
\bne 
  \frac{\tilde{t}_{n+1}^{\,(p)}}{\tilde{t}_{n}^{\,(p)}}-t
  =
  \frac{\sum_{i=1}^p
  \left( \! \begin{array}{c} p\\i\end{array} \! \right)
  (\frac{i}{p-i+1}-t) t_{n+i}+t_{n+p+1}-t\,t_n} {\sum_{i=0}^p \left
  ( \!\begin{array}{c} p\\i\end{array} \! \right) t_{n+i}} 
  \ .
  \label{expans} 
\ene
The goal is to prove that the latter expression goes to $0$ when
$p\rightarrow \infty$. 
To do so, we will split the sum in the numerator into four pieces
corresponding to the ranges
$$
   1 \le i < \sqrt{p} \:,\ 
   \sqrt{p} \le i < r_p-n_p \:,\ 
   r_p-n_p \le i < r_p + m_p \:,\ 
   r_p + m_p \le i \le p \:,\ 
$$
where
\bne
   r_p = {t \over 1+t} (p+1)  \label{rofp}
\ene
and where we leave the auxiliary sequences $n_p$ and $m_p$ free for the
moment, subject only to the conditions,
\bne
  n_p,m_p = o(p)=o(r_p) 
  \,, \qquad
  n_p,m_p\to\infty  \,{\rm as}\,  p \rightarrow \infty \,.
 \label{nm} 
\ene
We will also assume, without loss of generality, 
that $p$ is large enough so that $0 < \sqrt{p} < r_p-n_p$, whence we
will have
$$
  0 < \sqrt{p} < r_p-n_p < r_p+m_p < p \,.
$$
Observe here that $r_p$ has been chosen to make the expression
\bne
          {i \over p - i + 1} - t 
\ene
vanish when $i=r_p$.
Now, on performing our split in (\ref{expans}), we obtain
$$
  \frac {\tilde{t}_{n+1}^{\,(p)}} {\tilde{t}_{n}^{\,(p)}} - t
  =
  S_1 + S_2 + S_3 + S_4 \,,
$$
where 
(with `$\sum_{i=a}^{b}$' interpreted to imply $a\le{}i<b$ in the first
three cases)
we can write
\bea
0\leq -S_1 & = & 
\frac{\sum_{i=1}^{\sqrt{p}} \left( \! \begin{array}{c}
p\\i\end{array} \! \right) (t-\frac{i}{p-i+1}) t_{n+i}+tt_n}
{\sum_{i=0}^p \left( \!\begin{array}{c} p\\i\end{array} \! \right) t_{n+i}} 
\leq 
\frac{t\sum_{i=0}^{\sqrt{p}} \left( \!
\begin{array}{c} p\\i\end{array} \! \right) t_{n+i}}
{\sum_{i=0}^p
\left( \!\begin{array}{c} p\\i\end{array} \! \right) t_{n+i}}
= o(1) 
\label{S1} \\
0\leq -S_2 & = & \frac{\sum_{i=\sqrt{p}}^{r_p-n_p} \left( \! \begin{array}{c}
p\\i\end{array} \! \right) (t-\frac{i}{p-i+1}) t_{n+i}}{\sum_{i=0}^p
\left( \!\begin{array}{c} p\\i\end{array} \! \right) t_{n+i}} \nonumber\\ 
& \leq & 
\frac{\sum_{i=\sqrt{p}}^{r_p-n_p} 
\left( \! \begin{array}{c} p\\i\end{array} \! \right) 
 t^{i+1}(1+E_{\sqrt{p}})^{n+i}}
{\sum_{i=\sqrt{p}}^p
\left(\!\begin{array}{c} p\\i\end{array} \! \right) t^i(1-E_{\sqrt{p}})^{n+i}} 
\nonumber \\
& \sim &
\frac
 {\sum_{i=\sqrt{p}}^{r_p-n_p} 
 \left( \! \begin{array}{c} p\\i\end{array} \! \right) 
 t^{i+1}(1+E_{\sqrt{p}})^i}
{(1+t-tE_{\sqrt{p}})^p} \leq \frac{(1+E_{\sqrt{p}})^p}{(1+t-tE_{\sqrt{p}})^p} 
\left[ \sum_{i=\sqrt{p}}^{r_p-n_p} \left( \! \begin{array}{c}
p\\i\end{array} \! \right) t^{i+1}\right] =A_p 
\label{S2}\\
  |S_3| 
  & \leq & 
  \frac{\sum_{i=r_p-n_p}^{r_p+m_p} 
  {p \choose i}
  | \frac{i}{p-i+1}-t | 
  t_{n+i}}
  {\sum_{i=0}^p {p \choose i} t_{n+i}} 
  \leq 
  \sup_{i=r_p-n_p,r_p+m_p}(|\frac{i}{p-i+1}-t|) 
  = o(1) 
  \label{S3} \\
0 \leq S_4 
& = & 
\frac
  {\sum_{i=r_p+m_p}^p \left( \! \begin{array}{c}
   p\\i\end{array} \! \right) (\frac{i}{p-i+1}-t) t_{n+i}+t_{n+p+1}}
  {\sum_{i=0}^p \left( \!\begin{array}{c} p\\i\end{array} \! \right) t_{n+i}} 
\nonumber\\ 
& \le & 
\frac
  {\sum_{i=r_p+m_p}^{p+1} {p \choose i-1} t_{n+i}  }
  {\sum_{i=0}^p {p \choose i} t_{n+i}} 
\nonumber\\ 
& \leq & 
\frac
 {\sum_{i=r_p+m_p}^{p+1} \left( \! \begin{array}{c}
  p\\i-1 \end{array} \! \right) t^i(1+E_{\sqrt{p}})^{n+i}}
 {\sum_{i=\sqrt{p}}^p \left( \!\begin{array}{c} p\\i\end{array} \! \right) 
  t^i(1-E_{\sqrt{p}})^{n+i}} \nonumber\\
& \sim & 
\!\!\!\! \sum_{i=r_p+m_p-1}^p \!\!\!\! 
\frac{\left( \! \begin{array}{c} p\\i\end{array} \! \right) t^{i+1}
(1+E_{\sqrt{p}})^i}{(1+t-tE_{\sqrt{p}})^p} \leq \frac{(1+E_{\sqrt{p}})
^p}{(1+t-tE_{\sqrt{p}})^p} \left[ \sum_{i=r_p+m_p-1}^p \left( \!
\begin{array}{c} p\\i\end{array} \! \right) t^{i+1} \right] = B_p \,.
\label{S4} 
\ena
In these deductions, we used the Lemma in the final step of (\ref{S1}),
and we used it also 
(in the rather trivial special case,
$t_n\to[t(1-E_{\sqrt{p}})]^n$) to extend the sums in the denominators 
for $S_2$ and $S_4$ 
to the full range, $0\le{}i\le{}p$.
In (\ref{S3}), by narrowing the sum in its denominator, we converted the
second expression into a weighted mean of $|{i\over{p-i+1}}-t|$ with
the positive weights, ${p\choose{}i}t_{n+i}$ , thereby obtaining the third
expression, which is $o(1)$ thanks to (\ref{nm}).
Obviously, the only thing left to prove is 
that $S_2$ and $S_4$ go to zero as $p\rightarrow \infty$. 
For that, it will be sufficient to show that their bounds
$A_p$ and $B_p$ go to zero.

For future reference, we quote here the Stirling formula,
\be
     n! \sim \sqrt{2 \pi n} {\left({n \over e}\right)}^n 
\en
and its direct consequence (valid for $m,n\gg1$)
\bne
  \log{(m+n)! \over m! \; n!} =
  m \log{m+n\over m} + n \log{m+n\over n} + \half\log{m+n\over 2\pi mn}
  + o(1)
  \label{stirling2}
\ene

In order to bound $A_p$ and $B_p$, one can observe that the general
term of the sums appearing in them grows for $0\leq i\leq r_p$ and
decreases for $r_p \leq i \leq p$. This is because $r_p$ is
effectively defined such that the terms with $i=r_p$
and $i=r_p-1$ be equal. Then the sums can be bounded by their largest
term times the number of terms in the sum (which is itself smaller
than $p$). From there, the following bounds for $A_p$ and $B_p$ are
easily deduced:  
\be
A_p, B_p \leq \frac
 {p \left( \! \begin{array}{c} p \\ r_p+m \end{array} \! \right) 
   t^{r_p+m}   (1+E)^{p}}{(1+t-tE)^p}  
\ ,\en
where the simplified notations 
$$ m=-n_p+1, m_p \qquad E=E_{\sqrt{p}} $$
were introduced. This expression can be estimated with the aid of the
asymptotic formula (\ref{stirling2}). Substituting and expanding in
powers of small quantities like $m/p$ produces, 
after some tedium,
\be
 \ln(A_p, B_p) 
 \le 
 - \, {1+t\over 2} \, \frac{m^2}{r_p} 
 + \half\ln{p}  
 + 2r_p E
 + O(1) 
 + O(m^3/p^2)
 + O(r_p E^2)
\ ,
\en
which we will need only in the simplified form,
\bne
 \ln(A_p, B_p) 
 \le 
 {-(1+t)^2\over 2t} \frac{m^2}{p} 
 + O(\ln p)
 + O(p E)
 + O(m^3/p^2)  
 \ .
\label{ABform}
\ene

The only question now is whether
(consistently with (\ref{nm}))
we can choose 
the auxiliary sequence $m$ 
(representing either $m_p$ or $-n_p+1$)
so that 
the right hand side of (\ref{ABform})  diverges to $-\infty$,
i.e. whether we can choose $m=m(p)$
so that the first term in (\ref{ABform}) 
dominates the others for large $p$.
But this is not difficult.  For example, 
\be 
       m^2 = p^2 (E_{\sqrt{p}})^{1/2} + p^{3/2} 
\en 
meets all our requirements. With this choice, 
both $A_p$ and $B_p$ converge to zero as $p\rightarrow\infty$, 
and thus $S_2$ and $S_4$, which concludes the
proof that the sequence $\tilde{t}^{\,(p)}_{n+1}/\tilde{t}^{\,(p)}_{n}$
converges to $t$.

\subsection*{t=0}
This case can be treated in a manner very similar to the previous
case. So, suppose that 
\be
\lim_{n\rightarrow\infty} (t_n^{1/n})=0, 
\en
and let us introduce the auxiliary sequence 
\be
E_n=\sup_{i\geq n} (t_n^{1/n}) .
\en
To prove that the sequence $(t_n)$ flows to the fixed point $t_n=\delta_{1n}$ 
corresponding to $t=0$, 
it will be sufficient to prove 
(the somewhat stronger assertion)
that 
\bne
 \lim_{p\rightarrow +\infty} 
 \frac{\tilde{t}_{n+1}^{\,(p)}}
      {\tilde{t}_n ^{\,(p)}} 
 = 0
\label{goal0} 
\ene
for all $n\geq 1$. 

In writing (\ref{goal0}) this way, we have implicitly assumed that none
of the $\tilde{t}^{\,(p)}_n$ vanish, and this will hold automatically in
the generic case where the original $t_n$ are themselves all nonzero.
If, on the other hand, some of the $t_n$ do vanish, then there are two
possibilities.  Either the set of nonzero $t_n$ is infinite or finite.
If it is infinite, then, for any fixed $n$, only a finite number of the 
$\tilde{t}^{\,(p)}_n$ can vanish, so that the formulation (\ref{goal0})
remains valid as it stands, if we agree to omit a finite number of
initial values of the index $p$.   However, if all of the $t_n$ vanish
after some point, then our assertion must be reworded as follows.  Let 
$t_{n_0}$ be the last nonzero $t_n$ (recall that, by definition, not
{\it all} of the $t_n$ can vanish).  Then, for $p>n_0$,
$\tilde{t}^{\,(p)}_1,\tilde{t}^{\,(p)}_2,\cdots,\tilde{t}^{\,(p)}_{n_0}$
will all be $>0$, while $\tilde{t}^{\,(p)}_n$ will $=0$ for $n>n_0$.
Thus, our renormalized sequence $(\tilde{t}^{\,(p)}_n)$ will {\it
already} have converged to $\delta_{n1}$ for $n>n_0$, and we can
limit the assertion (\ref{goal0}) to $n<n_0$.
Having thus dealt with these special cases, we will assume henceforth
that all of the $\tilde{t}^{\,(p)}_n$ occurring in our discussion are
strictly positive.

Again, it is convenient to introduce an auxiliary sequence $m_p$, 
to be chosen later subject to the conditions
\bne 
    \sqrt{p} \leq m_p  = o(p)  \,.   \label{np} 
\ene
Then,
splitting the sum 
(\ref{D9-1}) 
in a manner similar to before, 
and applying similar techniques to bound the two resulting terms 
in ${\tilde{t}_{n+1}^{\,(p)}}/{\tilde{t}_n ^{\,(p)}}$,
we get
\bee
0
\leq
\frac{\tilde{t}_{n+1}^{\,(p)}}{\tilde{t}_n^{\,(p)}} 
& = & 
\frac{
\sum_{j=0}^{m_p} \left( \! \begin{array}{c} p\\j \end{array} \!
\right) t_{n+j}\frac{j}{p-j+1}+\sum_{i=m_p}^p \left( \!
\begin{array}{c} p\\i\end{array} \! \right) t_{n+i+1}}{\sum_{i=0}^p
\left( \! \begin{array}{c} p\\i\end{array} \! \right) t_{n+i}} \\ 
& \leq & 
\frac{m_p}{p-m_p+1} 
+ \frac{1}{t_n} \sum_{i=m_p}^p \left( \!
\begin{array}{c} p\\i\end{array} \! \right) t_{n+i+1}
\leq
o(1)+
 \frac{E_{\sqrt{p}}^{n+1}}{t_n} \sum_{i=m_p}^p \left( \!
\begin{array}{c} p\\i \end{array} \! \right) E_{\sqrt{p}}^i \ . 
\eee
Calling this last sum $A_p$, it will be sufficient to show that it
goes to zero as $p\rightarrow\infty$. 
As before, we can accomplish this, bounding $A_p$ by a geometric series.
Set 
\be
  u_i=\left( \! \begin{array}{c} p\\i \end{array} \! \right) E_{\sqrt{p}}^i 
 \ ,
\en
then 
\be
  \frac{u_{i+1}}{u_i}
 = \frac{p-i}{i+1}E_{\sqrt{p}} \leq \frac{p}{m_p} E_{\sqrt{p}} \ .
\en
In light of this result, a useful choice of $m_p$ is evidently
\bne
      m_p=\sup(p ({E_{\sqrt{p}}})^{1/2}, \sqrt{p})  \,;  \label{mpc}
\ene
for with this choice
$u_{i+1}/u_i{\le}\sqrt{E}{\to}0$ for $p\rightarrow \infty$, 
and the conditions (\ref{np}) are also satisfied. 
Than, by taking $p$ large enough, we can insure 
\be
  \frac{u_{i+1}}{u_i}\leq \frac{1}{2} \,,
\en
so that (omitting the subscripts on $m_p$ and $E_{\sqrt{p}}$ for brevity)
$$
  \sum\limits_{i=m}^p {p\choose i} E^i 
  = 
  \sum\limits_{i=m}^p u_i 
  \le 
  u_m \sum\limits_{i=0}^{p-m} 2^{-i} 
  \le 
  2 u_m 
  = 
  2 {p\choose m} E^m
$$
whence
$$
   A_p \le {2 \over t_n}  {p\choose m} E^{m+n+1} 
       \le {2 \over t_n}  {p\choose m} E^{m}
$$
or
$$
      \ln A_p \le O(1) + \ln {p\choose m} + m \ln E
$$
Invoking the Stirling formula (\ref{stirling2}) once again then yields,
after some simplification, 
$$
  \ln A_p 
  \,\le\, 
  O(1) + m \ln(pe/m) + m\ln{E}  
  \,\le\, 
  m \ln(me/p) + O(1)
$$
where the final step used 
that $\ln(E)\le2\ln(m/p)$  because of (\ref{mpc}).
This proves that $\ln(A_p)\to-\infty$ since  $m\ge\sqrt{p}\to+\infty$ 
and $m/p=o(1)\to0$ as $p\to\infty$.  In consequence,
$A_p\rightarrow 0$, which entails the desired convergence 
\be
 \lim_{p\rightarrow\infty} \left( \frac{\tilde{t}^{\,(p)}_{n+1}}{
 \tilde{t}^{\,(p)}_n} \right) =0 \, .
\en

\subsection*{t=$\infty$} 

In this case, the coefficients $t_n$ grow faster than any geometric
progression, and one might think, consistent with the $t\to\infty$
limit of (\ref{limproof}), that 
$\tilde{t}^{\,(p)}_{n+1}/\tilde{t}^{\,(p)}_n$
would tend to $+\infty$
after a large number of renormalizations.  Unfortunately this is not
necessarily so, but it seems clear, at least, that it would follow 
under the stronger hypothesis that not only the $(t_n)^{1/n}$, 
but the ratios $\rho_n=t_{n+1}/t_n$ converged to $+\infty$ with $n$.
In such a case, the renormalized dynamics would be trivial in the sense
that it would produce only chains, as one can see from
(\ref{D8-1}) and (\ref{D8-2}) above
(because the unique $\lambda(\vpi,m)$ with $\vpi=n$ 
would swamp all others in the
limit $p\to\infty$).

\subsection*{Multiple limits}
When the sequence $t_n^{1/n}$ does not converge, it is difficult to
conclude anything in general.
However, if, 
for a given integer $q$, 
the $q$ subsequences $t_{k(q+1)+i}$ 
all converge as $k\rightarrow\infty$,
for $0\leq i \leq q$:
\bne
 \lim_{k\rightarrow\infty} t_{k(q+1)+i}^\frac{1}{k(q+1)+i}=T_i,
 \label{ansatzmulti} 
\ene
then the $p$-times renormalized sequence $t_n^{(p)}$ 
tends as $p\to\infty$ toward
the fixed point corresponding to $t=\sup (T_i)$. 
This is because,
as we are going to show,
after $q$ renormalizations
one has
\bne
  \lim_{n\rightarrow\infty}((\tilde{t}^{(q)}_n)^{1/n})=\sup_{0\leq i\leq q}
   (T_i)=t \,,    \label{corol} 
\ene
so that the previous result applies. We can restrict ourselves to the
case when $t\not=0$, because otherwise, since $T_i\geq 0$
automatically from the positivity of $t_n$, we would have $T_i=0$ for
all $0\leq i\leq q$ and therefore $t_n^{1/n}$ itself would converge to
$0$, a case treated in the previous subsection.

To show (\ref{corol}), we define the auxiliary sequence 
\bne
u_n=\sup_{n\leq i \leq n+q}(t_i),\label{defu} 
\ene
so that 
\bea
 (t^{(q)}_n)^{1/n}
 =\left( \sum_{j=0}^q \left( \begin{array}{c} q \\ j \end{array} \right) 
 t_{n+j} \right)^{1/n} 
 & \leq & u_n^{1/n} 2^{q/n}  
 \ \sim \ u_n^{1/n} \ ,   \nonumber \\
 & \geq & u_n^{1/n} \left( \begin{array}{c} q \\ q/2 \end{array} \right)^{1/n} 
   \sim u_n^{1/n} \,. 
\label{ineqtu} 
\ena
The latter inequality is because at least one of the terms $t_{n+j}$ in
the sum is equal to $u_n$ and the 
binomial coefficient 
$\left( \begin{array}{c} q \\ q/2 \end{array} \right)$ 
is the largest of the lot. 
The inequalities (\ref{ineqtu}) show in particular that 
\bne 
  (t_n^{(q)})^{1/n}\sim u_n^{1/n}.\label{tsimu} 
\ene

Now, to study the convergence of the sequence $u_n^{1/n}$, we split it into
$q+1$ subsequences with indices $k(q+1)+i$, $0\leq i \leq q$, which  cover
the whole sequence. Thus, reordering the elements of the supremum in
(\ref{defu}), we have 
\be 
u_{k(q+1)+i}^{1/(k(q+1)+i)}=\sup_{0\leq j\leq q} [t_{(k+H(i-j))(q+1)
+j}^{1/(k(q+1)+i)}],
\en
where $H(p)$ is an Heaviside--like function defined by 
\bee
H(n) & = & 1 \mbox{ if } x>0 \\
& = & 0 \mbox{ otherwise.}
\eee
Then from the ansatz (\ref{ansatzmulti}), 
\be 
t_{(k+H(i-j))(q+1)+j}^{1/(k(q+1)+i)}=(t_{(k+H(i-j))(q+1)+j}^{1/[(k+H(i
-j))(q+1)+j]})^{[(k+H(i-j))(q+1)+j]/[k(q+1)+i]} \sim T_j, 
\en 
and therefore the subsequence $u_{k(q+1)+i}$ converges: 
\be
\lim_{k\rightarrow\infty} (u_{k(q+1)+i}^{1/(k(q+1)+i)})=\sup_{0\leq
j\leq q} (T_j)=t.
\en
Since each of the subsequences converges to the same limit $t$ and
since 
they cover the whole sequence $u_n$, we deduce  
\be
 \lim_{n\rightarrow\infty}u_n^{1/n}=t \,,
\en
and hence the result (\ref{corol}) from eq. (\ref{tsimu}).

In more general cases, it seems to be impossible to decide the
question of convergence with our present methods.  However, since the
sequences we have not studied will have converging subsequences in
stretches of diverging size, it seems that, as the number $p$ of
renormalizations goes to infinity, longer and longer initial stretches
of the renormalized $t_n$ will, for almost all $p$, look like
percolation for some $t$ (which will in general vary with $p$).
However, this same idea also gives a way to construct sequences which
will not flow to any single fixed point, but alternate between various
ones, and we describe such a counterexample in the next section.

\section{A counterexample and a conjecture}
\label{counterex}
We have seen that a large number of sequences $(t_n)$ yield
trajectories $(\tilde{t}^{\,(p)}_n)$ which converge to fixed points of
$\P$, including all those $(t_n)$ for which
$\lim\limits_{n\to\infty}t_n^{\,1/n}$ exists.  In this section, we
will describe a counterexample to the supposition that all
trajectories whatsoever approach limits.  However, our counterexample
does not contradict the weaker supposition that every trajectory, in
some sense ``spends most of its time near to the fixed point set as a
whole'', and we conclude this section with a sample conjecture to that
effect.

Now let $(t_n)$ be any starting sequence of $t$'s.  If
the corresponding trajectory $(\tilde{t}^{\,(p)}_n)$ of renormalized
$t$'s converged to some limit $(\tilde{t}^{\,(\infty)}_n)$ then in
particular the ratio 
$\tilde{t}^{\,(p)}_2/\tilde{t}^{\,(p)}_1\ideq f(p)$
would have to converge to 
$\tilde{t}^{\,(\infty)}_2/\tilde{t}^{\,(\infty)}_1$.
To construct our counterexample, then, 
it suffices to contrive $(t_n)$ so that $f(p)$ has no limit.  
But we claim that, 
if $a_1,a_2,a_3\cdots$ is any sequence whatsoever of positive reals, 
and if $\epsilon>0$ is arbitrary,
then we can find a starting sequence $(t_n)$
and a subsidiary sequence of integers $p_1<p_2<p_3\cdots$ such that 
$|f(p_n)-a_n|<\epsilon$ for all $n$.  For example, let us try
\bee
   t_n&=& (a_1)^n  \ \ {\rm  for} \ \ 1     \le n \le p_1+2 \,, \cr
   t_n&=& (a_2)^n  \ \ {\rm  for} \ \ p_1+2  <  n \le p_2+2 \,, \cr
   t_n&=& (a_3)^n  \ \ {\rm  for} \ \ p_2+2  <  n \le p_3+2 \,, \cr
   &&\mathrm{etcetera.}        \cr
\eee
If, for any $k$, we were to send $p_k$ to infinity, then 
obviously $\lim\limits_{n\to\infty}(t_n)^{1/n}$ would be $a_k$
and our main result above would assure us that 
$f(n)\stackrel{n\to\infty}{\longrightarrow}a_k$;
hence we can  certainly find $p_k$ great enough that 
$|f(p_k)-a_k|<\epsilon$.
Moreover, we can subsequently alter the values of $t_n$ for $n>p_k+2$
without affecting $f(p_k)$, since the latter clearly depends only on
$\tilde{t}^{\,(p_k)}_1$ and $\tilde{t}^{\,(p_k)}_2$,
and these in turn depend only on $t_n$ for 
$n\le{}p_k+1$ and $n\le{}p_k+2$, respectively.
Hence, we can always select the $p_k$ to verify our claim. Figure 1
illustrates this technique with a closely related example in which 
$\tilde{t}^{\,(p)}_2 / \tilde{t}^{\,(p)}_1$ oscillates 
between (near to) 1/2 and (near to) 2,  
the specific choice of $(t_n)$ in that case being:
$t_n=n$ if $n$ is a power of 2, and $t_n=0$ otherwise,
i.e. $(t_n)=(1,2,0,4,0,0,0,8,0,0,0,\cdots)$.

\begin{figure}[htbp]
\center
\scalebox{1.3}{\includegraphics{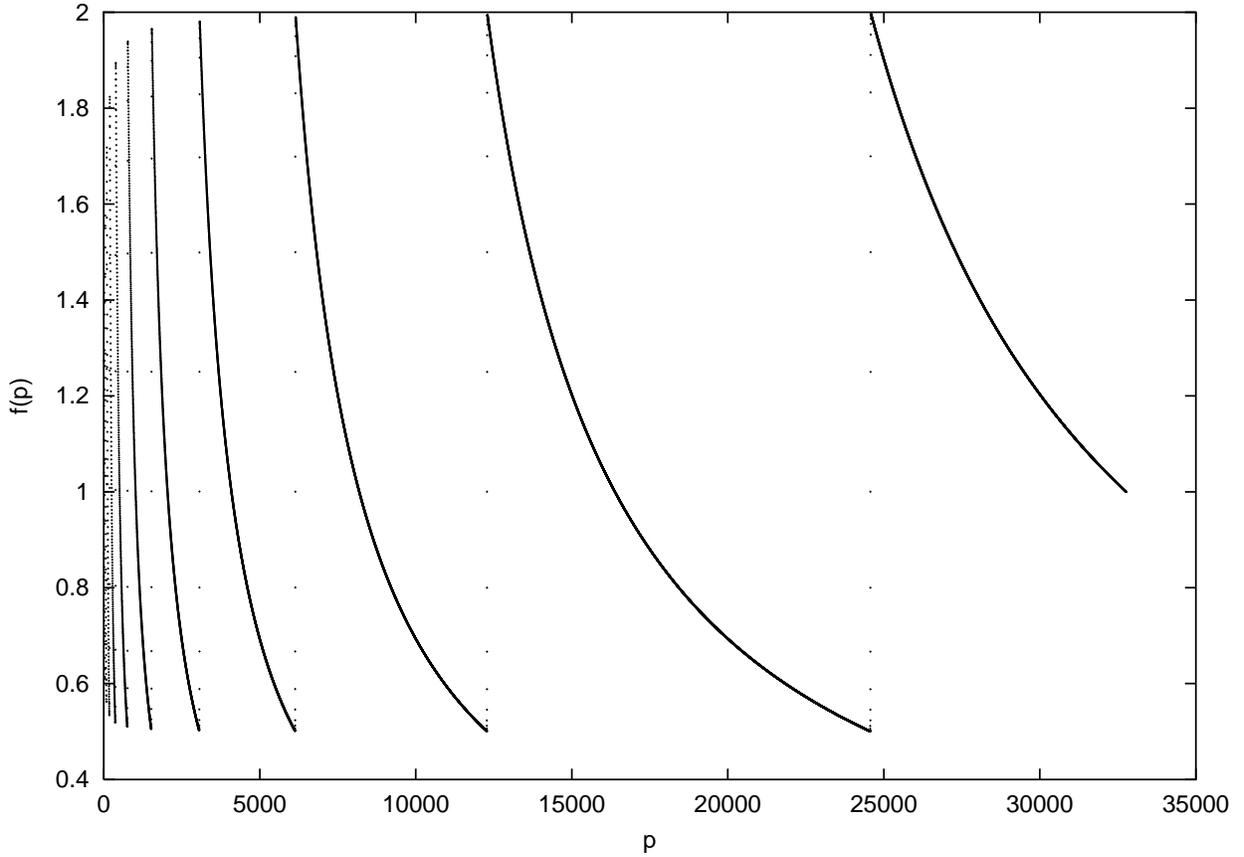}}
\caption{Flow of the ratio $\tilde{t}^{\,(p)}_2/\tilde{t}^{\,(p)}_1$
under renormalization}
\end{figure}

Closer inspection of this example, and in particular of Figure 1,
reveals that the $\tilde{t}^{\,(p)}_n$ are not behaving completely
chaotically.  Rather, the precipitous jumps are narrowly localized in
``time'', while between them, the ratios $\rho_n=t_{n+1}/t_n$ vary only
gradually.  Together with the evidence from other examples, this
suggests that the ``moments'' when the renormalized $t_n$ deviate from
percolation-like values are few and far between.  In some, yet to be
specified sense, then, $(t_n)$ would be spending most of the time near
the ``percolation submanifold'', with but brief excursions to other
regions.  In order to render this idea somewhat more definite, let us
cast it in the form of a conjecture.  For brevity, let us say that a
real number $r$ is ``within $\eps$ of $\infty$'' when $r>1/\eps$.  Then
we conjecture that:
$(\forall \eps>0)$ $(\forall m)$ (the fraction of $n<N$ for which the
$m$ ratios
$$
     \tilde{t}^{\,(n)}_2/\tilde{t}^{\,(n)}_1, \ 
     \tilde{t}^{\,(n)}_3/\tilde{t}^{\,(n)}_2, \ 
     \cdots,                             
     \tilde{t}^{\,(n)}_{m+1}/\tilde{t}^{\,(n)}_m
$$
are all within $\eps$ of each other (or all within $\eps$ of $\infty$)
tends to unity as $N\to\infty$).

\section{Further reflections}
The main theorem proved above guarantees pointwise convergence to
percolation dynamics when $\lim\limits_{n\to\infty}(t_n)^{1/n}$
exists.  However the topology of pointwise convergence is rather
coarse, and using it, we could not even claim, for example, that the
``basin of attraction'' of the fixed point set was extensive enough to
include any open neighborhood of the latter (because no open set in
this topology can control more than a finite number of terms of the
sequence $(t_n)$).  On the contrary, the set of sequences {\it not}
converging to the fixed-point set would be dense in the space ${\cal
T}$ of all sequences.  Thus this topology does not seem to provide a
useful language for discussing the global features of our ``RG flow''.

{}From a physical point of view, pointwise convergence provides
information with limited temporal validity.  For example, pointwise
convergence to $t_n=0$ means that, following a cosmic cycle comprising
a large number of causet elements, the ensuing expansion will be
tree-like for a long, but in general finite period.  To guarantee
permanent tree-like behavior, one would need something like uniform
convergence of the $t_n$, or perhaps more appropriately, uniform
convergence of their ratios, $t_{n+1}/t_n$.  An interesting question,
therefore, is whether, by suitably strengthening its hypotheses, one
could prove an analog of our main theorem for the topology of uniform
convergence.  Another natural extension of the present work, possibly
of greater urgency, would be to explore, not just {\it which}
trajectories approach the fixed point set, but also the {\it manner}
in which they approach it; for this could help answer the question of
how general is the phenomenon discovered in \cite{dou} according to
which the ``cosmic big number'' associated with the so-called
``flatness problem'' could be explained by the hypothesis that the
universe has undergone several previous cycles of expansion and
collapse.

Finally, let us remark that the question of which topology (or
topologies) is most suited to discuss the ``renormalization group flow''
with which we have been concerned in this paper is inseparable from the
wider question of which sequences $(t_n)$ represent dynamical laws that
are genuinely conceivable from the physical point of view.  One can
imagine, for example, uncontrollably divergent sequences, and if
${\cal{T}}$ is really the set of {\it all} sequences, then one might
expect its dimensionality as a projective space to exceed $\aleph_0$,
the dimension of a separable Hilbert space.  In that case, the normal
tools of functional analysis would seem to be unavailing.  

But does it really make sense that $t_n$ could grow arbitrarily
rapidly (or, for that matter, be a number whose precision required for
its expression arbitrarily many significant digits)?  Our mathematical
framework allows this, but only --- it would seem --- as an artifact
of the procedure by which one introduces the dynamical laws in a
``non-material'' manner, as if ``from outside the universe''.  It
would be more satisfactory if the laws could somehow be understood as
embodied in the structure of material universe --- in this case in the
structure of the causal set.  But then, the number of possible laws
should itself be limited at any stage of the growth of the causal set,
meaning that the number of possibilities for $t_0,\cdots,t_n$ (the
parameters that determine the dynamics at stage $n$) would be bounded
by something like the number of possibilities for a causet of $n$
elements, a number which grows only as $2^{O(n^2)}$.  It might be,
therefore, that the space in which our renormalization transformation
$\P$ acts is in reality not just of countable dimension but actually
of countable cardinality.  Important as this would be for the deeper
understanding of the questions studied in this paper, it would be
premature at present to speculate on how such a limitation on
dynamical laws might work out in detail someday.  Here, we wished only
to raise the possibility, and in doing so to call to mind the idea
that kinematics must ultimately fuse with (or absorb) dynamics as part
of the further progress of fundamental physical theory.


\bigskip\noindent

R.D.S would like to thank the Aspen Center for Physics, where parts of
this paper were written.

This work was supported by a joint NSF--CONACyT grant number E120.0462/2000.

X. Martin was supported by CONACyT and SNI--M\'exico, and D. O'Connor by the CONACyT grant 30422-E.

This research was partly supported by NSF grants PHY-9600620 and
INT-9908763 and by a grant from the Office of Research and Computing
of Syracuse University.

\bigskip
\vspace{3.5mm}
\noindent{\bf \Large Appendix }\par\nobreak\smallskip\noindent
\vspace{1.5mm}
In this appendix, we derive an identity used in Section
\ref{formulation} of the main text, namely
\bne
   \lambda(\vpi,m \,|\, \tilde{t}^{\,(N)}) = \lambda(\vpi+N,m \,|\, t)
\label{A1}
\ene
where
\bne
  \lambda(\vpi,m \,|\, t) = \sum\limits_k {\vpi-m \choose k-m} t_k 
  \label{A2}
\ene
and
\bne
   \tilde{t}^{\,(N)}_n = \sum\limits_j {N \choose j} t_{n+j}  \,.
\ene
The proof relies on a second (well known) identity,
\bne
  \sum\limits_{i+j=k} {m \choose i} {n \choose j} = {m+n \choose k} \,,
  \label{A4}
\ene
which itself follows from expanding out a third (trivial) identity,
\be
   (1+x)^m (1+x)^n = (1+x)^{m+n} \,.
\ee
{}From (\ref{A2})-(\ref{A4}), we obtain
\bee
\lambda(\vpi,m \,|\, \tilde{t}^{\,(N)}) &=&
\sum\limits_k{\vpi-m \choose k-m} \tilde{t}^{\,(N)}_k
\cr
&=&
\sum\limits_k{\vpi-m \choose k-m} \sum\limits_j {N\choose j} t_{k+j}
\cr
&=&
  \sum\limits_k \sum\limits_j  {\vpi-m \choose k-m}{N\choose j}  t_{k+j} 
\cr
&=&
\sum\limits_l  
  \left(\sum\limits_{j+k=l} {\vpi-m \choose k-m} {N\choose j}\right)  t_l
\cr
&=&
\sum\limits_l  {\vpi-m + N \choose l-m}  t_l
=
\lambda(\vpi+N,m \,|\, t)  \,,
\eee
which is (\ref{A1}).

\end{document}